# Ordering and dimensional crossovers in metallic glasses and liquids

*David Z. Chen[1,*], Qi An[2], William A. Goddard III[2], Julia R. Greer[1,3]*

[1]Division of Engineering and Applied Sciences,

California Institute of Technology, Pasadena CA, 91125, USA

[2]Materials and Process Simulation Center,

California Institute of Technology, Pasadena CA, 91125, USA

[3]The Kavli Nanoscience Institute, California Institute of Technology, Pasadena CA, 91125, USA

*Corresponding author, Email: dzchen@caltech.edu

Abstract: The atomic-level structures of liquids and glasses are amorphous, lacking long-range order. We characterize the atomic structures by integrating radial distribution functions (RDF) from molecular dynamics (MD) simulations for several metallic liquids and glasses: $Cu_{46}Zr_{54}$, $Ni_{80}Al_{20}$, $Ni_{33.3}Zr_{66.7}$, and $Pd_{82}Si_{18}$. Resulting cumulative coordination numbers (CN) show that metallic liquids have a dimension of d = 2.55 ± 0.06 from the center atom to the first coordination shell and metallic glasses have d = 2.71 ± 0.04, both less than 3. Between the first and second coordination shells, both phases crossover to a dimension of d = 3, as for a crystal. Observations from discrete atom center-of-mass position counting are corroborated by continuously counting Cu glass- and liquid-phase atoms on an artificial grid, which accounts for the occupied atomic volume. Results from Cu grid analysis show short-range d = 2.65 for Cu liquid and d = 2.76 for Cu glass. Cu grid structures crossover to d = 3 at ξ~8 Å (~3 atomic diameters). We study the evolution of local structural dimensions during quenching and discuss its correlation with the glass transition phenomenon.

# I. INTRODUCTION

The viscosities and relaxation times of glasses and liquids across the glass transition temperature ($T_g$) are separated by many orders of magnitude.[1] This large increase in viscosity over a short temperature range is not accompanied by significant changes in the long-range atomic structure, which remains amorphous. Metallic glasses are locally more ordered in the short- and medium-range than their liquid counterparts,[2,3] but this ordering plays an ambiguous role in the glass transition.[4] A structural model that captures both liquids and glasses is useful for understanding the amorphous structure and the subtle changes, if any, that occur across $T_g$ and their potential connection to the glass transition phenomenon.

The local dimension, d, describes how, on average, the mass of atoms within a spherical section of material with radius r scales, $M(r) \propto r^d$.[5] In relating the positions of the first sharp X-ray diffraction peaks ($q_1$) to sample volume (V), several groups have reported a scaling relationship in metallic glasses, with exponent, d~2.31-2.5, which deviates from the d = 3 expected under the assumption that $q_1 \propto 1/a$, where a is the interatomic spacing.[6-9] Recent real-space measurements on $Ti_{62}Cu_{38}$ also revealed a dimensionality of roughly 2.5.[10] Experiments on electrostatically levitated metallic liquids also show a non-cubic power law exponent of d~2.28,[11] albeit with a limited range in data and a significant amount of scatter.[12] Without translational symmetry, the connection between diffraction peak positions and interatomic distances in amorphous materials is not simple.[12] Nonetheless, the estimated power law exponents are related to the local dimension of the atomic structure, and observations of an exponent/dimension less than 3 have led to suggestions of an underlying fractal structure in metallic glasses.[6,8] However, the long-range scaling relationship in metallic glass structure is not fractal over all length scales because macroscopic pores or voids are absent in their microstructure, and such pores are a defining characteristic of fractals that maintain their scaling relationships over long ranges (e.g. the Sierpinski triangle).

In response to this inconsistency, Chen et al. proposed that metallic glasses at the atomic-level can be described using percolation,[8] a model that captures the interconnectivity of sites on a lattice or spheres in a continuum.[5] Three-dimensional percolation models, such as hard sphere and overlapping sphere continuum models, exhibit a fractal dimension of d~2.52 at lengths below a correlation length, $\xi$, and a crossover to a dimension d~3 above $\xi$, where $\xi$ is roughly the diameter/length of finite, non-percolating clusters.[5] Using molecular dynamics (MD) simulations, Chen et al. found that two distinct metallic glasses have short-range dimensions of d~2.5 below ξ~2 atomic diameters and a dimension of 3 occurs over longer lengths. This suggested that metallic glasses are structurally similar to a continuum percolation (i.e. spatially-random coalescence) of spherical particles.[8] This crossover at ξ may explain the anomalous non-cubic scaling exponents in $q_1$ vs. V observed experimentally in macroscopically homogeneous and fully dense metallic glasses and liquids.[6-8, 11] Such a connection between percolation structure and glasses has also been suggested by Orbach, who applied percolation theory to describe high frequency (short wavelength) vibrational states in glassy systems and also suggested that amorphous materials may exhibit fractal properties at short length scales.[13]

The question remains whether liquids exhibit a crossover in dimension from d < 3 to d = 3. Percolation structure has been studied in hard spheres,[14, 15] overlapping spheres,[16, 17] and recently metallic glasses,[8] suggesting a possible connection to metallic liquids, which share structural similarities with both metallic glasses and hard sphere systems[18]. It would be interesting to study the development of this ordering as a function of temperature, across the glass transition. One previous method to measure dimension utilized hydrostatic pressures to induce peak shifts in radial distribution functions (RDF) that were compared to corresponding volume changes.[8] However, this pressure-induced peak shift method is not well suited for studying liquids, in which atomic rearrangement and exchange of neighbors leads to significant structural changes under pressure. The correlation lengths, ξ, can only be inferred based on the scaling of various peaks. Moreover, the broadness of the RDF peaks leads to results that are sensitive to the specific method of generating and measuring the RDF.[19] To overcome these challenges, we integrated the RDFs to obtain cumulative coordination numbers (CN). This integral method estimates the local dimension of the structure without the need for applying hydrostatic pressures or measuring small shifts in broad amorphous peak positions, which are methods that were used previously.[8] With this CN analysis, we observe a crossover in dimension from d = 2.55 ± 0.06 in metallic liquids and d = 2.71 ± 0.04 in metallic glasses, to d = 3 for the second coordination shell and beyond, suggesting that ξ~3 atomic diameters.

## II. DIMENSION AND CROSSOVER

One measure of dimension comes from the scaling of extensive properties with size such as mass, i.e. $M(r) \propto r^d$, where $M(r)$ is the mass contained in a sphere of radius r. $M(r)$ is calculated as an average over the entire system by choosing different atoms as the center of the sphere.[5] In our analysis, we use the value CN+1 to represent the average number of atoms within a sphere of radius r (1 added to account for the center atom), an extensive property that is proportional to average mass. In percolation, the scaling relationship for a system above the percolation threshold, $\phi_c$, exhibits a crossover in dimension from d~2.52 to d~3 at ξ, where $\xi \propto (\phi - \phi_c)^{-\nu}$.[5] The parameter definitions are: $\phi$ is the packing fraction, $\nu = 0.8764$ is the critical exponent for the correlation length,[20] and $\phi_c$ is the percolation threshold in 3-dimensional continuum percolation.[5] From percolation theory, the expected crossover point for several of the metallic systems studied here has been roughly estimated to be ξ~2.[8] This value represents the average size of non-percolating clusters in units of atomic diameters, and suggests that the crossover occurs around the first atomic coordination shell. To avoid inaccuracies that may arise from determining precise peak shifts in broad amorphous peaks, we obtain the dimension of each atomic structure by measuring $d(\ln(CN_{grid}))/d(\ln(r))$ for $Cu_{46}Zr_{54}$, $Ni_{80}Al_{20}$, $Ni_{33.3}Zr_{66.7}$, and $Pd_{82}Si_{18}$ metallic liquids and glasses. We find that a crossover from d < 3 to d = 3 occurs in all cases beyond the first to second coordination shell. We compare these results to those for pure Cu and Zr (SI) in liquid and crystalline phases.

### A. Metallic glasses

We measure d by performing a linear fit between the radius of the center atom, $r_{avg}$, and the outer radius of the first coordination shell, $r_{1s}$. The $r_{avg}$ is defined as the average radii of the atoms in the binary systems (i.e. for $Cu_{46}Zr_{54}$, $r_{avg} = 0.46r_{Cu} + 0.54r_{Zr}$, refer to SI). There is on average one atom (i.e. the center atom) within this radius, making it an appropriate first point in the analysis of the dimension. Using this approach, we establish the following estimates of dimensions: d = 2.68 for $Ni_{80}Al_{20}$, d = 2.73 for $Ni_{33.3}Zr_{66.7}$, d = 2.66 for $Pd_{82}Si_{18}$, and d = 2.74 or 2.73 for $Cu_{46}Zr_{54}$ using $FF_1$[21] or $FF_2$[22], respectively (Figure 2), all at 300 K. The average dimension for metallic glasses of d = 2.71 ± 0.04 is ~0.19 higher than what would be expected from percolation theory, where d~2.52,[5] and is higher than previous estimates of ~2.3-2.5[6,7] (diffraction experiments) and ~2.5[8] (molecular dynamics with hydrostatic pressure). In the region between the center atom and first coordination shell, $r_{avg}$-$r_{1s}$, CN rises sharply due to the discrete nature of the atom counting procedure, which bins the atoms according to their center of mass position, providing no information on their physical volume (i.e. from excluded volume interaction) and precluding the counting of fractions of atoms. A continuous measure of the CN that captures this missing structural information might give a smooth, filled-in curve between the center atom and first coordination shell and a more accurate estimate of short-range dimension (refer to Section C).  Between the outer radii of the first and second coordination shells, $r_{1s}$-$r_{2s}$, the dimension crosses over to 3 for all cases, suggesting that these metallic glasses have a correlation length of ξ~3.56 atom diameters, which is slightly higher than previous estimates.[8] Here ξ is estimated using $(r_{1s}+r_{2s})/2r_{avg}$.[23] Within the first to second coordination shell, free volume arising from packing inefficiencies contributes to a reduced dimensionality in the structure. This reduced (< 3) dimension cannot proliferate to greater lengths because the free volume necessarily remains smaller than the volume occupied by atoms, whose relative positions are dictated by long-range attraction and low kinetic energy. At longer length scales, where free volume is less significant and the atom clusters appear closely packed, we find that the dimension of the structure is 3.

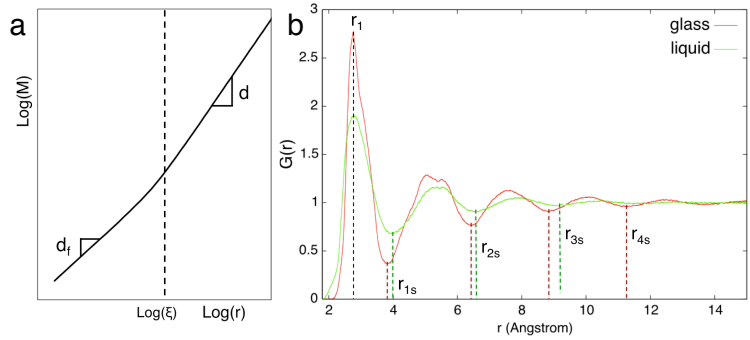

**Figure 1.** a) Diagram of expected crossover in log-log plot of mass versus radius. Short-range fractal dimension $d_f$ crosses over to long-range dimension d at the correlation length ξ. b) Radial distribution functions for $Cu_{46}Zr_{54}$ ($FF_2$) in the glass and liquid phase. Dashed lines indicate positions for the first peak, $r_1$, and coordination shells, $r_{is}$, where i = 1-4.

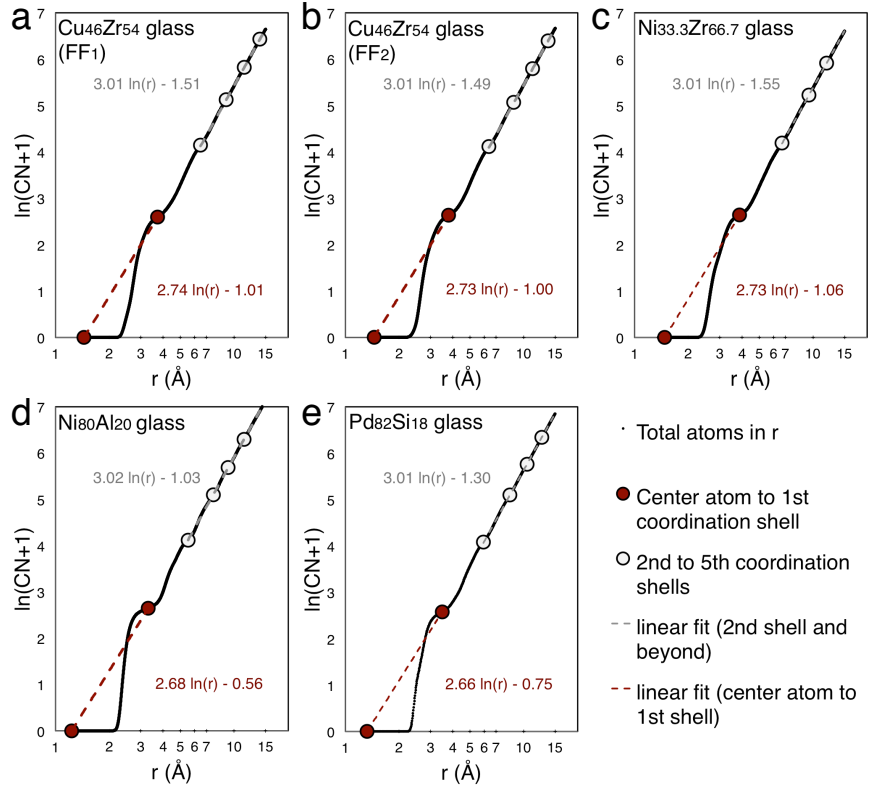

**Figure 2.** Log-log plots of total atom number (CN+1) versus radius, r, showing local dimension in metallic glasses of $Cu_{46}Zr_{54}$ a) $FF_1$, b) $FF_2$, c) $Ni_{80}Al_{20}$, d) $Ni_{33.3}Zr_{66.7}$, and e) $Pd_{82}Si_{18}$. Short-range dimension, d = 2.71 ± 0.04, is measured through a linear fit between the radius of the center atom and the outer radius of the first coordination shell. Long-range dimension, d = 3, is measured from a linear fit of points beyond the outer radius of the second coordination shell.

B. Metallic liquids

Applying the same method to metallic liquids, we measure d = 2.57 for $Cu_{46}Zr_{54}$ $FF_1$ at 2500 K, d = 2.55 for $FF_2$ at 2000 K, d = 2.48 for $Ni_{80}Al_{20}$ at 3000 K, d = 2.64 for $Ni_{33.3}Zr_{66.7}$ at 2500 K, and d = 2.53 for $Pd_{82}Si_{18}$ at 2000 K (Figure 3) from $r_{avg}$ to $r_{1s}$. These estimates are dependent on temperature, as the position of $r_{1s}$ changes due to thermal expansion (see section D). The average value of d = 2.55 ± 0.06 is in line with the value of ~2.52 from percolation theory,[5] and is ~0.16 lower than the average value in our metallic glasses. This difference in local dimensions in liquid and glassy phases may be related to the accumulation of dense ordered clusters, such as icosahedra, across the glass transition, which pack more efficiently and reduce local free volume.[3, 24, 25] A crossover in dimension from d < 3 to d = 3 occurs in roughly the same region as in the metallic glasses, which suggests that the liquids are structurally analogous to percolation structures with a correlation length of ξ~3.68 atomic diameters, slightly higher than previous suggestions.[8] In percolation theory, the correlation length is inversely related to the atomic packing fraction ($\xi \propto (\phi - \phi_c)^{-\nu}$), and more loosely packed liquid structures may exhibit longer ξ. Metallic liquids are dense, possessing packing fractions of around ϕ~0.67 ($FF_2$ at 2000 K), a value that is only ~8% lower than their glassy counterparts (ϕ~0.73 for $FF_2$ glass at 300 K). To observe structures with ξ~4 diameters or longer, we estimate that we would need to study liquids and glasses with packing fractions in the neighborhood of ϕ~0.5, which is not feasible for our metallic systems, as a first-order phase transition to the gaseous phase would likely precede such a low packing fraction in the liquid phase.

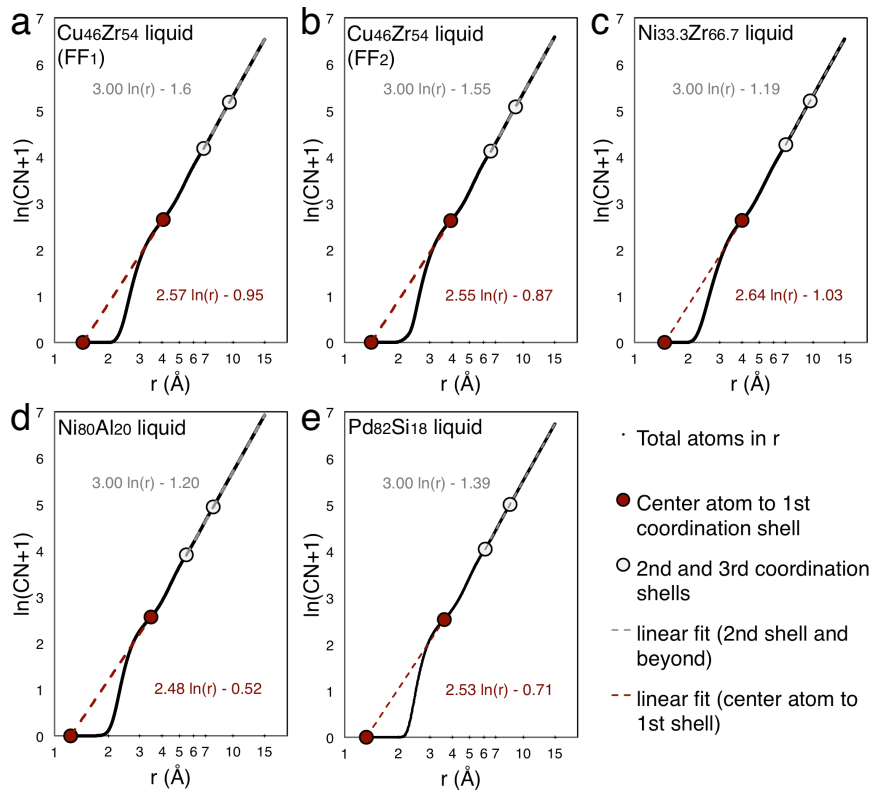

**Figure 3.** Log-log plots of total atom number (CN+1) versus radius, r, showing local dimension for metallic liquids of $Cu_{46}Zr_{54}$ a) $FF_1$ at 2500 K, b) $FF_2$ at 2000 K, c) $Ni_{80}Al_{20}$ at 3000 K, d) $Ni_{33.3}Zr_{66.7}$ at 2500 K, and e) $Pd_{82}Si_{18}$ at 2000 K. Short-range dimension, d = 2.55 ± 0.06, is measured through linear fit between the radius of the center atom and the outer radius of the first

coordination shell. Long-range dimension, d = 3, is measured from a linear fit of points beyond the outer radius of the second coordination shell.

### C. Comparison to Copper and grid analysis

We compare our results to those for crystalline Cu at 300 K, which has a dimension of 2.93 between the center atom and the minimum after the first peak (measured at the midpoint between the first and second peak).[23] Beyond the first peak, the dimension is ~3 (Figure 4a). We expect the long-range crystal dimension to be exactly 3 owing to its close-packed cubic structure. In the short-range, the crystal dimension should be slightly less than 3, owing to finite-temperature fluctuations and presence of defects.

Comparison of the crystalline (300 K), glassy (300 K), and liquid (2500 K) phases of Cu shows that the major contribution to < 3 dimensionality in the liquid and glassy phases is the short-range structure, which, due to fluctuations in free volume, can be locally more open. The overall coordination number curve is shifted toward higher radii for the liquid phase, which reduces its short-range dimension. The short-range structure in the glass phase appears denser and more ordered compared to the liquid – the coordination number rises more steeply in the first shell, increasing d towards a close-packed, crystalline value.

The discrete nature of our atom-counting procedure introduces error into the estimates for local dimension and makes the measurements of short-range dimension in these structures delicate, as the fitting is performed over only two points. This motivates a method to count the atoms continuously by modeling them as spheres that occupy a volume based on their atomic radii. For this purpose, we represented our Cu system as points on a grid, which occupy the physical volume of each Cu atom with a 0.3-Å resolution (Figure 4b). To generate the grid, we impose a mesh onto the entire system with a specified spacing. We select a grid spacing of 0.3 Å in order to optimize spatial resolution while weighing computation time. We keep the nodes on the mesh that lie within $r_{Cu}$ of the center of mass of each Cu atom, where $r_{Cu}$ is the radius of Cu, ~1.28 Å, and we reject nodes that do not meet this criterion. The remaining nodes are the grid points that occupy the physical space of our Cu atoms. To perform the atom counting, we take the partial RDFs of each atom center of mass position with respect to the grid points and normalize by the average number of grids per atom.

With the grid method, we find that the short-range dimension of the Cu liquid is 2.65 and that of the Cu glass is 2.76 (Figure 4c). We estimate $\xi \sim 3$ using $\xi = r_c/r_{Cu}$, where $r_c \sim 4$ Å and $r_{Cu} = 1.278$ Å, for both phases. The correlation length is slightly longer for the liquid phase than the glass phase, owing to a lower global density, which leads to a longer $r_c$. This lower global density (higher local free volume concentration) also contributes to a lower short-range dimension in the Cu liquid compared to the glass, ~2.65 vs. 2.76. This effect dominates over the averaging effect due to temperature fluctuations, which may serve to increase local dimension (see section D about $d_v$). Our observations on the relative dimensions from the grid method of counting CN corroborate those from the two-point analysis involving $r_{avg}$ and $r_{1s}$. The estimates for local dimension are more accurate in the grid analysis, as the fitting is performed over a longer range of r, rather than two points. Even so, it is likely more pertinent to compare the relative dimensions of identical systems under various conditions rather than consider their absolute values. This is evident from the observation that the local dimensionality is r-dependent (see secondary-axis plot in Figure 4c). The local $1^{st}$ derivatives of the $\ln(CN_{grid})$ plots, $d(\ln(CN_{grid}))/d(\ln(r))$ vs. r, show that the local dimensions of these systems vary depending on the real-space region of the structure. In the Cu glass, the local dimension is close to 3 in a narrow peak between $r_{avg}$ (1.278 Å) and $r_1$ (~2.5 Å). In the Cu liquid, this peak is lower and shifted toward longer r. Interestingly, there is a stable real-space region between $r_1$ and $r_2$ where the < 3 dimensionality reliably occurs for both glassy and liquid phases.

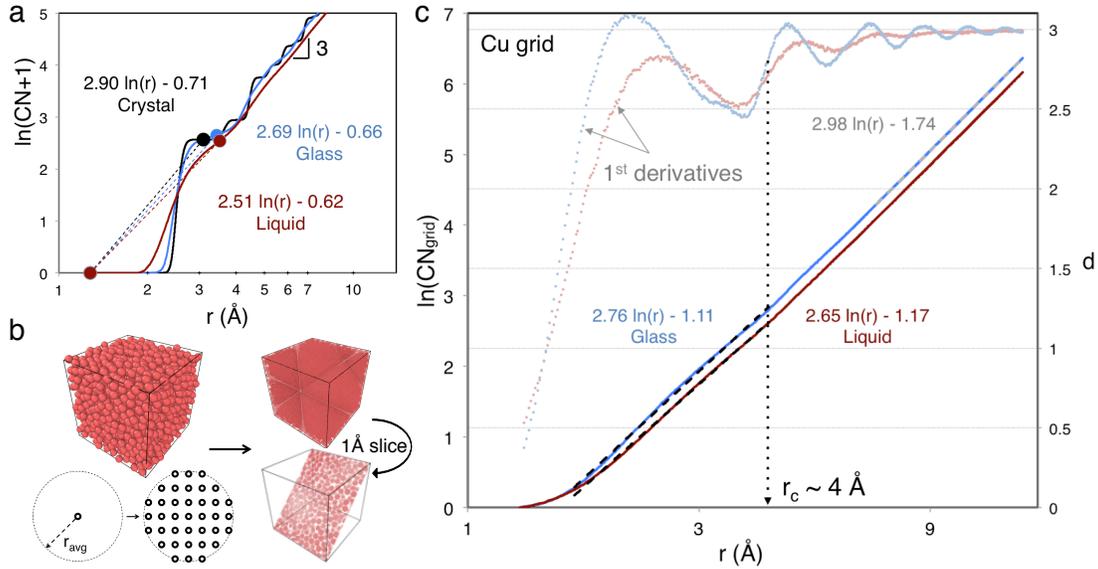

**Figure 4.** Comparison of crossovers in pure Cu systems using discrete and grid counting methods. a) d~2.90 in Cu crystal (300 K), d~2.69 in Cu glass (300 K), and d~2.51 in Cu liquid (2500 K) below ξ with CN counted by atom center positions. b) Schematic of the grid procedure. Cu atoms in the simulation box (left) are replaced by effective grid points representing their physical volume. Grid points capture the overall atomic structure (see 1 Å slice, right). c) Crossovers in dimension from d~2.65 and d~2.76 to d~3 for Cu liquid and glass, respectively using a grid method for continuous counting. Here $CN_{grid}$ is the normalized coordination number based on counting grids within each atom. Secondary axis (right side): $d(\ln(CN_{grid}))/d(\ln(r))$ versus r showing a distinct crossover near ξ~8 Å.

### D. Temperature effects on atomic structure during quench

We examine the evolution of local dimensions within real-space regions of interest in our Cu systems as a function of temperature during quenching from the liquid state to the glassy state (Figure 5). Each temperature snapshot is taken via quenching from the immediately higher temperature. The short-range dimension, $d_s$, which we define heuristically as ranging from ~$1.2r_{Cu}$ to ~$3.2r_{Cu}$, increases roughly linearly with decreasing temperatures. This is somewhat unexpected, as the global volume change during cooling is linear in the liquid and glassy regions, while strictly nonlinear near the glass transition.[23] The short-range dimensional changes indicated by $d_s$ do not reflect the same trend as that from the global volume, showing instead a lack of an inflection point near $T_g$. This suggests that $d_s$ is mostly temperature-dependent, and is not sensitive to the glass transition. It also suggests that the local dimension in a real-space segment within $d_s$ must be decreasing very rapidly near $T_g$. The valley in $d(\ln(CN_{grid}))/d(\ln(r))$ versus r from $\sim r_1$ to $\sim 3.2 r_{Cu}$, corresponding roughly to the center of the first nearest neighbor to the edge of the second nearest neighbor, has a local dimension, denoted $d_v$, that is very sensitive to the glass transition. $d_v$ hovers around 2.6 at temperatures above 1500 K, and dips abruptly below 2.55 on cooling past 1200 K, close to the glass transition temperature of $T_g \sim 1150$ K. This abrupt shift corresponds also to the appearance of a shoulder in the first minimum, which indicates the development of ordered clusters. The $d_v$ region relates to the amount of free volume in the system around the first neighbor. The liquid phase has a higher $d_v$ due to stronger thermal fluctuations occurring at higher temperatures, which play an averaging role on the local dimension. A reduction in $d_v$ indicates increased local free volume just beyond the nearest neighbor, which suggests, somewhat counter-intuitively, increased local order via ordered clusters. This is analogous to the development of interstitial volume, which greatly decreases local density within a narrow region in r during crystallization.

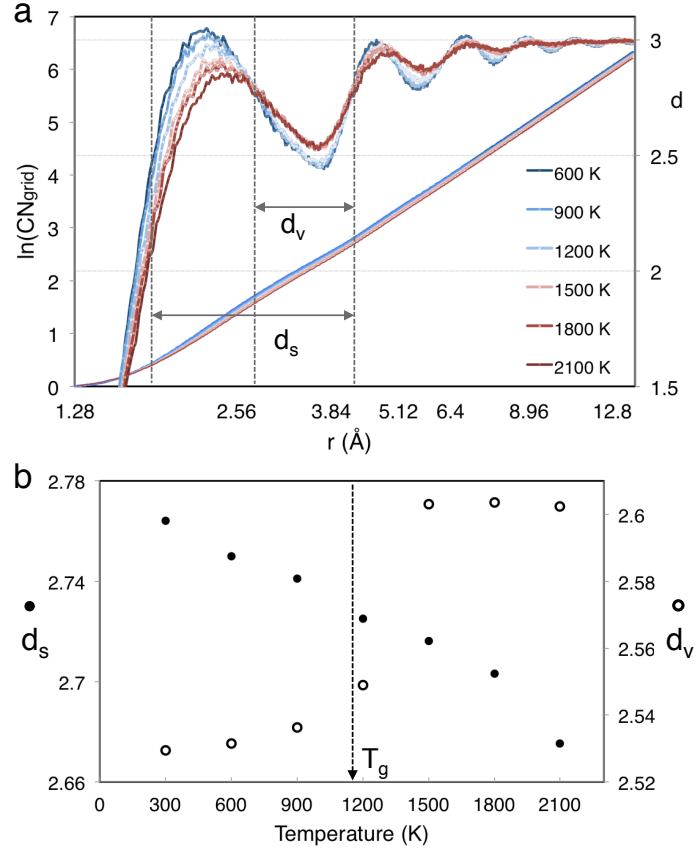

**Figure 5.** Temperature effects on local dimensions in Cu glass and liquids using grid counting. a) $\ln(CN_{grid})$ versus r and $d(\ln(CN_{grid}))/d(\ln(r))$ versus r plots at temperatures from 600-2100 K at 300 K intervals. We define heuristically intervals for short-range dimension, $d_s$, ~1.2—3.2$r_{Cu}$, and valley dimension, $d_v$, ~2—3.2$r_{Cu}$, where the local dimension is largely temperature insensitive but appears to be sensitive to the liquid/glass phase. $r_{Cu}$ = 1.28 Å. b) $d_s$ and $d_v$ versus temperature; $d_s$ increases roughly linearly as temperature is decreased, and $d_v$ appears sensitive to the glass transition ($T_g$ ~ 1150 K).

We consider the evolution of $d_v$ as a function of volume fraction, $\phi$, and the effects of global volume change on the local dimensionality (Figure 6). We calculate $\phi$ using $N_{grid}V_{grid}/V_s$, where $N_{grid}$ is the number of occupied grids in the system, $V_{grid}$ is the volume of each grid voxel, and $V_s$ is the total system volume. We observe an inflection point in $d_v$ versus $\phi$ around $\phi$~0.64-0.66. Notably, this value corresponds to the random close packed (RCP) value and the maximally random jammed (MRJ) value in monodisperse hard spheres.[26, 27] This inflection point also occurs close to the packing fraction at the glass transition, $\phi \sim 0.66$. Other methods of calculating $\phi$, such as taking $\phi = N_{Cu}V_{Cu}/V_s$, where $N_{Cu}$ and $V_{Cu}$ are, respectively, the number and volume of Cu atoms, yield similar results.

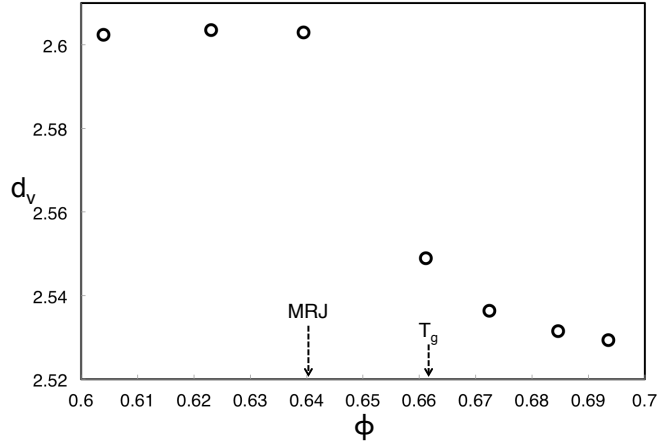

**Figure 6.** $d_v$ versus $\phi$ showing a connection between the maximally random jammed (MRJ, $\phi \sim$ 0.64) state and onset of $T_g$.

## III. DISCUSSION

The glass transition may be related to the densification/ordering that occurs in the local glass structure, but the connection is not clear. Previous analyses comparing amorphous and crystalline structures have emphasized that radii ratios of ~0.6-0.95 in binary systems favors formation of amorphous phases,[28] and local icosahedral structure in the first shell plays an important role in driving glass formation for Cu-Zr-Al metallic glasses.[21,29] In our analysis, we observe an increase in d from ~2.55-2.65 to ~2.71-2.76 from the liquid to glass phases, suggesting that some ordering occurs across the glass transition in these metallic alloys and metals. This ordering can be seen more clearly in the grid analysis of Cu liquid and glass structures (Figure 4c), where the two main observations are: 1) the short-range dimension, $d(\ln(CN_{grid}))/d(\ln(r))$ vs. r plot from ~1.5 Å to ~4 Å, is $d_{Cuglass} \approx 2.76$ for the Cu glass, ~0.11 higher than that of the liquid phase, which has $d_{Culiquid} \approx 2.65$, and 2) $d(\ln(CN_{grid}))/d(\ln(r))$ vs. r shows sharpening in the first peak of the Cu glass, reaching a slope of around 3, indicating ordering in the first nearest neighbors, and a shoulder appears near the first minimum, indicating the development of ordered clusters.

Absolute changes in d, ~0.11-0.16, across the glass transition are small, representing only a ~4-6% increase. However, keeping in mind that the values for d are roughly constrained to be from 2 to 3, as these structures occupy 3-dimensional space, the relative changes in slopes are actually closer to ~20-30%.

The liquid-glass transition appears to be a universal phenomenon in that any liquid can vitrify with sufficiently fast cooling.[4, 30] Diverging relaxation time and viscosity can happen with or without accompanying structural changes. For example, symptoms of the glass transition such as the jump in heat capacity and logarithmic increase of $T_g$ with quench rate can be explained without invoking phase transitions and thermodynamics, by considering that the systems stop relaxing within the experimental timescale.[31] In these metallic systems, the structural changes that appear across the glass transition may be unique – other common glasses such as covalent network glasses or molecular glasses have not yet been studied in this way, although the methods presented here can be extended to study those systems. Nonetheless, the structural effects observed in this study on metallic glasses may be instructive for a more general understanding of the liquid-glass transition (refer to SI for additional discussion).

The short-range dimension in our metallic glasses, d~2.71-2.76, in contrast to the metallic liquids, deviates considerably from percolation models, where the fractal dimension is ~2.52. In simple percolation models, the constituent units occupy lattice sites or are allowed to overlap one another[5] such that no limit exists for the site occupancy probability or volume fraction of overlapped spheres. In real systems and hard sphere percolation models, the constituent spherical particles (e.g. metallic atoms) have excluded volume. This gives rise to fundamental limits in the random close packing fraction of hard spheres, which is $\phi$~0.637 for monodisperse spheres,[32] and ~0.64-0.83 for bi-disperse spheres, depending on their radii ratios and compositions.[33] Stable binary metallic glasses, while not perfectly represented by hard spheres, have high packing fractions: ~0.73 for our $Cu_{46}Zr_{54}$ ($FF_2$) and above ~0.7 for other binary alloys.[34] Interestingly, our monatomic Cu system exhibits sensitive changes in $d_v$ near $T_g$ and at a volume packing fraction of $\phi$~0.64. This corresponds closely with RCP and MRJ states in monodisperse hard spheres. The densification/ordering that occurs in these systems at the atomic level may be due to the frustration and jamming of the atoms, which approach and exceed the maximal packing fractions allowed by the random packing of spheres, arresting molecular motion. A similar idea has been explored in granular materials; for example, Xia, et al. found that polytetrahedra serve as structural elements to glassy order in hard-sphere particle glasses, forming a globally jammed fractal structure.[35] The mechanism for geometrical constraint in our systems may be similar to ideas in jamming or rigidity percolation.[36, 37]

## IV. SUMMARY

We find that the cumulative CN analysis shows a crossover in dimension for both metallic glasses and liquids. We observe that the short-range dimension is less than 3, d~2.55-2.71 for both liquids and glasses using two methods: 1) two-point analysis from linear fit between $r_{avg}$ and $r_{1s}$ in binary systems, and 2) grid analysis of continuously counting grid points representing monatomic Cu systems. The long-range dimension crosses over to 3 beyond the first coordination shell. Analysis of the structural evolution during quenching suggests that ordering develops across the glass transition as short-range dimension increases roughly linearly with decreasing temperatures. Observations of local dimensions between ~2—3.2$r_{Cu}$ in Cu shows sensitivity to the glass transition and a correlation with the packing fraction around RCP and MRJ states, suggesting that densification during cooling of metallic liquids may be arrested by fundamental packing limits near the glass transition.

## V. ACKNOWLEDGMENTS

The authors would like to acknowledge Jun Ding and Mark Asta for pointing out the sensitivity in measuring precise RDF peak positions. The authors gratefully acknowledge the financial support of the US Department of Energy, Office of the Basic Energy Sciences (DOE-BES) under grant DE-SC0006599 and NASA's Space Technology Research Grants Program through J.R.G's Early Career grants. Parts of the computations were carried out on the SHC computers (Caltech Center for Advanced Computing Research) provided by the Department of Energy National Nuclear Security Administration PSAAP project at Caltech (DE-FC52-08NA28613) and by the NSF DMR-0520565 CSEM computer cluster. Q.A. and W.A.G. received support from NSF (DMR-1436985). This material is based upon work supported by the National Science Foundation (NSF) Graduate Research Fellowship under Grant No. DGE-1144469. Any opinion, findings, and conclusions or recommendations expressed in the material are those of the authors and do not necessarily reflect the views of the NSF.

## VI. APPENDIX: MOLECULAR DYNAMICS METHODS

All molecular dynamics simulations of the metallic liquids and glasses discussed here used embedded atom model (EAM) potentials:

- The $Cu_{46}Zr_{54}$ systems (54,000 atoms) were prepared using two potentials, Cheng et al.[21] ($FF_1$) and Mendelev et al.[22] ($FF_2$).

- The $Ni_{80}Al_{20}$ systems (32,000 atoms) were prepared using Pun et al.,[38]

- The $Ni_{33.3}Zr_{66.7}$ systems (32,000 atoms) were prepared using Mendelev et al.[39], and

- The $Pd_{82}Si_{18}$ systems (32,000) were prepared using Ding et al[40].

Cutoff distance: $FF_1$ – 6.5 Å, $FF_2$ – 7.6 Å, NiAl – 6.3 Å, NiZr – 7.6 Å, Cu – 7.6 Å, PdSi – 6.5 Å. Cooling procedure: cooled from melt to room temperature over 1000000 ps (steps of 0.001 ps). Thermalization at the end of cooling: fixed NPT at 300 K and 0 Pa for 100000 ps

We selected four binary metallic glasses and liquids: $Cu_{46}Zr_{54}$, $Ni_{80}Al_{20}$, $Ni_{33.3}Zr_{66.7}$, and $Pd_{82}Si_{18}$. Among these four MGs, Cu-Zr, Ni-Zr and Pd-Si belong to metal-metal MGs and Pd-Si belongs to metal-metalloid MGs. The binary Cu-Zr and Pd-Si MGs have been synthesized in experiments.[41] Although bulk metallic glasses have not been formed in binary Ni-Zr and Ni-Al systems, they are interesting to study in simulations because they have good (simulated) glass forming ability.[42, 43]

In all cases the binary metallic glasses were quenched from the liquid phase (2000-3000 K) at a rate of $\sim 10^{12}$ K/s to room temperature (300 K). The Cu crystal (13,500 atoms), liquid (2048 atoms) and glass (2048 atoms) are prepared from $FF_2$. The Cu metallic glass was quenched at a rate of $\sim 10^{14}$ K/s.

For the grid analysis, we first mapped the whole space onto grid sites on a cubic lattice with spacing ~0.3 Å. We remove grid points outside the average radius of the atoms by marking all of the grid points within one atomic radius from an atomic center. The remaining grid points fill the excluded volume of our systems. The total number of grid points is 657545 for the Cu glass and 642106 for the Cu liquid.

The RDFs were calculated by binning the atomic structure (100,000 bins for binary systems and Cu crystal, 5000 bins for Cu liquid and glass). Coordination numbers are obtained by integrating the total RDF. The $CN_{grid}$ value is taken from the partial RDF from the Cu atom center positions to the grid points. We normalize the final $CN_{grid}$ value by the average number of grid points within each atom ($CN_{grid}$ at $r = r_{avg}$).

Supporting Information: coordination number dimension analysis for Zr crystal, different RDF binning conditions, and applied hydrostatic pressures (30 GPa).

amorphous systems. Precise measurements of pressure-induced RDF peak shifts on the order of ~0.01 Å are necessary to derive an accurate estimate of d. Hydrostatic pressure-induced peak shifts are small (~0.01 Å/GPa), making the overall fit for d particularly sensitive to the position of the initial peak at 0 Gpa. In the previous analysis, a change in initial peak position of ~0.01 Å can shift the exponent for d by ~0.2. In our current analysis using cumulative CN, A ~0.01 Å change in rcenter or $r_{1s}$ would result in a ~0.02 shift in the estimate for d. With our new analysis, the previous estimates of d~2.5 for $Cu_{46}Zr_{54}$ $FF_1$ and $FF_2$ and $Ni_4Al$ now become d~2.74, 2.73, 2.68, respectively.